\documentclass[namedreferences]{solarphysics}
\usepackage[optionalrh]{spr-sola-addons} % For Solar Physics
\usepackage{graphicx}
\usepackage{color}
\usepackage{url}             % For breaking URLs easily trough lines
\newcommand{\aap}{    {\it Astron. Astrophys.}}

\newcommand{\apj}{    {\it Astrophys. J.}}
\newcommand{\apjl}{   {\it Astrophys. J. Lett.}}
\newcommand{\apss}{   {\it Astrophys. Spa. Sci.}}

\newcommand{\jgr}{    {\it J. Geophys. Res.}}

\newcommand{\pasj}{   {\it Publ. Astron. Soc. Japan}}

\newcommand{\solphys}{{\it Solar Phys.}}

\newcommand{\ssr}{    {\it Space Sci. Rev.}}

\ifx \doiurl    \undefined \def
\doiurl#1{\href{http://dx.doi.org/#1}{\textsf{DOI}}}\fi \ifx \adsurl
\undefined \def
\adsurl#1{\href{http://adsabs.harvard.edu/abs/#1}{\textsf{ADS}}}\fi
\ifx \arxivurl  \undefined \def
\arxivurl#1{\href{http://arxiv.org/abs/#1}{\textsf{arXiv}}}\fi

\begin{document}
\begin{article}
\begin{opening}

\title{The 26 December 2001 Solar Eruptive Event Responsible for GLE63.
II.~Multi-Loop Structure of Microwave Sources in a Major
Long-Duration flare}

\author{\inits{V.V.}\fnm{V.}~\lnm{Grechnev}\orcid{0000-0001-5308-6336}}
\author{A.M.~\surname{Uralov}\sep
        V.I.~\surname{Kiselev}\sep
        A.A.~\surname{Kochanov}}

 \runningauthor{V.V. Grechnev \textit{et al.}}
 \runningtitle{The 26 December 2001 Solar Flare}

 \institute{Institute of Solar-Terrestrial Physics SB RAS,
                  Lermontov St.\ 126A, Irkutsk 664033, Russia
                  email: \url{grechnev@iszf.irk.ru}
                  email: \url{uralov@iszf.irk.ru}
                  email: \url{valentin_kiselev@iszf.irk.ru}
                  email: \url{kochanov@iszf.irk.ru}}

\date{Received ; accepted }

\begin{abstract}
Analysis of the observations of the SOL2001-12-26 event related to
ground-level-event GLE63, including microwave spectra and images
from the \textit{Nobeyama Radioheliograph} at 17 and 34\,GHz,
\textit{Siberian Solar Radio Telescope} at 5.7\,GHz, and
\textit{Transition Region and Coronal Explorer} in 1600\,\AA\ has
led to the following results. A flare ribbon overlapped with the
sunspot umbra, which is typical of large particle events. Atypical
were: i)~long duration of the flare of more than one hour;
ii)~moderate intensity of a microwave burst, about $10^4$\,sfu;
iii)~low peak frequency of the gyrosynchrotron spectrum, around
6\,GHz; and its insensitivity to the flux increase by more than
one order of magnitude. This was accompanied by a nearly constant
ratio of the flux emitted by the volume in the high-frequency part
of the spectrum to its elevated low-frequency part determined by
the area of the source. With the self-similarity of the spectrum,
a similarity was observed between the moving microwave sources and
the brightest parts of the flare ribbons in 1600\,\AA. Comparison
of the 17\,GHz and 1600\,\AA\ images has confirmed that the
microwave sources were associated with multiple flare loops, whose
footpoints appeared in ultraviolet as intermittent bright kernels.
To understand the properties of the event, we simulated its
microwave emission using a system of several homogeneous
gyrosynchrotron sources above the ribbons. The scatter between the
spectra and sizes of the individual sources is determined by the
inhomogeneity of the magnetic field within the ribbons. The
microwave flux is mainly governed by the magnetic flux passing
through the ribbons and the sources. An apparent simplicity of
microwave structures is caused by a poorer spatial resolution and
dynamic range of the microwave imaging. The results indicate that
microwave manifestations of accelerated electrons correspond to
the structures observed in thermal emissions, as well-known models
predict.

\end{abstract}

\end{opening}

\section{Introduction}
\label{S-introduction}

We continue a study (\citealp{GrechnevKochanov2016}; Article~I) on
the 26 December 2001 event (SOL2001-12-26). This solar
eruptive-flare event produced a strong flux of solar energetic
particles (SEPs, mainly protons) near Earth and a 63rd
ground-level enhancement of cosmic-ray intensity (GLE63). GLEs
represent the highest-energy extremity of SEPs (see,
\textit{e.g.}, \citealp{Cliver2006, Nitta2012}; and references
therein). GLEs are rare events; seventy-two GLEs only have been
recorded since 1942. The rare occurrence of GLEs hampers
understanding their origins and finding consistent patterns that
might govern their appearance and properties.

Unlike electrons, whose signatures are manifold throughout the
whole electromagnetic range, accelerated protons and heavier ions
can only be detected on the Sun from nuclear $\gamma$-ray emission
lines appearing in their interactions with dense material (see,
\textit{e.g.}, \citealp{Vilmer2011}). Solar $\gamma$-ray
observations have been very limited in the past. No $\gamma$-ray
images were available before 2002. Due to the observational
limitations, considerations of the origins of SEPs and,
especially, GLEs mainly refer to the hypotheses proposed several
years ago (see, \textit{e.g.}, \citealp{Kallenrode2003,
Aschwanden2012} for a review). On the other hand, studies based on
recent observations \citep{Cheng2011, Zimovets2012, Grechnev2013a,
Grechnev2014, Grechnev2015b, Grechnev2016} indicate that some of
these hypotheses might need refinement.

GLEs are typically produced by major eruptive events associated
with big flares (mostly of the GOES X-class), fast coronal mass
ejections (CMEs), and strong microwave bursts. The common
association with different solar energetic phenomena
\citep{Kahler1982, Dierckxsens2015, Trottet2015} hampers the
identification of the origins of SEPs and GLEs. All of these
circumstances show how important the analysis of a solar source of
each GLE is.

With a general correspondence between the magnitudes of SEPs, CME
speeds, and flare manifestations, there are few outliers from the
correlations \citep{Grechnev2013a, Grechnev2015a}. Strong fluxes
of high-energy protons were observed near Earth after these events
associated with moderate microwave bursts. The correlation between
the proton fluences, on the one hand, and microwave and soft X-ray
(SXR) fluences, on the other hand, is considerably higher. This is
difficult to understand, if SEPs are exceptionally
shock-accelerated. A flare-related contribution could also be
significant in the events, whose proton productivity was enhanced
for some unclear reasons. One of these events was the
SOL2001-12-26 eruptive-flare event responsible for GLE63.

This solar event has not been comprehensively studied because of
its limited observations. We are not aware of either low-coronal
observations of an eruption or hard X-ray (HXR) data. On the other
hand, the event was observed in microwaves by the \textit{Siberian
Solar Radio Telescope} (SSRT: \citealp{Smolkov1986,
Grechnev2003ssrt}) at 5.7\,GHz; the \textit{Nobeyama
Radioheliograph} (NoRH: \citealp{Nakajima1994}) at 17 and 34\,GHz,
and in ultraviolet (UV) by the \textit{Transition Region and
Coronal Explorer} (TRACE: \citealp{Handy1999}).

The SSRT observations have been studied in Article~I. Its
conclusions are:

\begin{enumerate}

\item Most likely, GLE63 was caused by the M7.1 solar event in
active region (AR) 9742 (N08\,W54). Contribution from a concurrent
far-side event is unlikely.

\item The flare was much longer than other GLE-related flares and
consisted of two parts. The first, possibly eruptive, flare and a
moderate microwave burst started at 04:30 and reached an M1.6
level. The main flare, up to M7.1, with a much stronger burst
started at 05:04, when a CME was launched.

\item The main flare involved strong magnetic fields presumably
associated with a sunspot in the west part of AR~9742.

\item Two non-thermal sources observed at 5.7\,GHz initially
approached each other nearly along the magnetic neutral line, and
then moved away from it like expanding ribbons, as if they were
associated with the legs of the flare arcade.

\end{enumerate}

It was difficult to confirm and expand these indications in
Article~I based solely on the SSRT data because of their
insufficient spatial resolution and coalignment accuracy. To
verify and elaborate these results, observations of the flare
arcade or ribbons in a different spectral range, where they are
clearly visible, should be compared with microwave data of a
higher spatial resolution.

Of special interest is a conjectured localization of the
non-thermal microwave sources in the legs of the flare arcade.
This possibility does not contradict a commonly accepted view on
the flare process; however, HXR and microwaves almost always show
a few simple non-thermal sources. Ribbon-like HXR structures have
been observed in exceptional events \citep{MasudaKosugiHudson2001,
LiuLeeGaryWang2007}. The simplicity and confinement of non-thermal
sources in impulsive flares suggested involvement of one to two
loops \citep{Hanaoka1996, Hanaoka1997, Nishio1997,
GrechnevNakajima2002}. Later observational studies extended this
view up to some long-duration flares
\citep{TzatzakisNindosAlissandrakis2008}. A concept of a single
microwave-emitting loop became dominant.

One of the major purposes of our companion articles is to find the
possible causes of the contrast between the rich proton outcome of
this solar event and its moderate manifestations in microwaves. We
consider \textit{a priori} the contributions from the acceleration
processes both in the flare and by the shock wave to be possible
\citep{Grechnev2015a}. Accordingly, we examine the 26 December
2001 flare in this article (Article~II) and eruptive event in
Article~III (Grechnev \textit{et al.}, 2017, in preparation). Here
we endeavor to shed further light on the listed issues related to
the flare itself and its microwave emission, analyzing the
observations carried out by TRACE in 1600\,\AA\ and by NoRH at 17
and 34\,GHz.

%%%%%%%%%%%%%%%%%%%%%%%%%%%%%%%%%%%%%%%%%%%%%%%%%%%

\section{Observations}

\subsection{Parts of the Flare}
\label{S-flare}

TRACE observed the flare mostly in the 1600\,\AA\ channel, which
we use. The 1600\,\AA\ images are similar to those observed in the
H$\alpha$ line, but they do not show filaments. Some images were
produced less often in other UV channels, 1550 and 1700\,\AA, and
in white light. The whole set that we analyze consists of 668
images observed in 1600\,\AA\ from 04:23 to 05:22 (all times
hereafter refer to UTC). The imaging interval between 63 images
from 04:23 to 04:55 was around 30 seconds, and then it decreased
to about 2 seconds from 05:04 to 05:22. The exposure time was 0.43
seconds before 04:30, then 0.26 seconds until 05:00, and it
changed cyclically between 0.020, 0.031, and 0.052 seconds during
the main flare. A \textsf{2001-12-26\_TRACE\_1600A.mpg} movie in
the Electronic Supplementary Material presents the whole flare
observed in 1600\,\AA.

We processed the TRACE images as follows: The background level was
found for each image as a highest-probability value in its
histogram. This individual level was subtracted from each image,
which was converted to a reference exposure time of 0.43 seconds.
The offset between the 1600\,\AA\ and white-light channels was
corrected from their coalignment. Comparison with white-light
images produced by the \textit{Michelson Doppler Imager} (MDI:
\citealp{Scherrer1995}) onboard SOHO revealed a roll angle of
$0.9^{\circ}$ in the TRACE images, which was compensated, and
refined their absolute pointing coordinates. All of the images
were transformed to 05:04 to compensate for the solar rotation.
Bright defects produced by high-energy particles were reduced in
calculations by using a minimum of two images separated by a small
interval \citep{Grechnev2004}. Remaining issues of the processing
described here are insignificant.

\begin{figure} % {1}
  \centerline{\includegraphics[width=\textwidth]
   {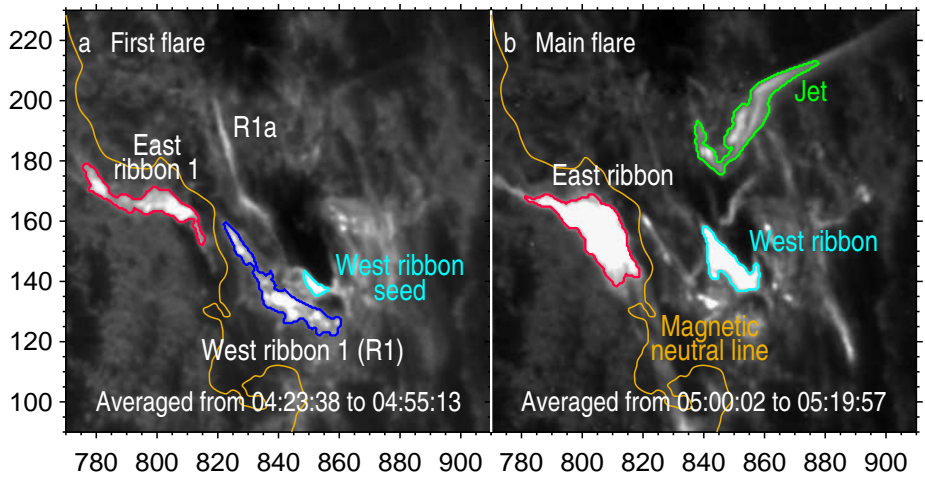}
   }
  \caption{Averaged TRACE 1600\,\AA\ images showing the regions
involved in the first flare (a) and main flare (b). The orange
contour traces the magnetic neutral line at the photospheric
level. a)~The ribbons in the first flare (red and blue contours).
The light-blue contour outlines a seed of the west ribbon, which
evolved in the major flare. b)~The ribbons in the major flare (red
and light-blue contours). The green contour outlines a jet and a
part of its base. The axes show hereafter the coordinates in
arcsec from the solar disk center.}
  \label{F-flare_regions}
  \end{figure}

The major flare configurations are presented by the TRACE images
averaged during each of the two flare parts in Figures
\ref{F-flare_regions}a and \ref{F-flare_regions}b. The images are
shown within a limited brightness range and scaled by a power-law
function with a $\gamma$ of 0.7. A dark region near the centers of
the panels corresponds to a sunspot. The orange contour traces the
major magnetic-polarity inversion (neutral) line at the
photospheric level calculated from a SOHO/MDI magnetogram observed
on 26 December at 04:51.

The averaged images and movie present complex flare
configurations. The first flare started from the appearance of two
extended thin ribbons at both sides of the neutral line (a remote
northeast brightening is beyond the field of view of
Figure~\ref{F-flare_regions}). The east ribbon evolved in the
first flare within the red contour. Three structures are
magnetically conjugate to the east ribbon. A long thin ribbon R1a
was active early in the event, and then it smoothly faded. We do
not consider this region. Intermittent brightenings within the
dark-blue contour were active during the first flare, and then
became less important. A small bright region started to grow near
the sunspot (light-blue contour).

 \begin{figure} % {2}
  \centerline{\includegraphics[width=0.70\textwidth]
   {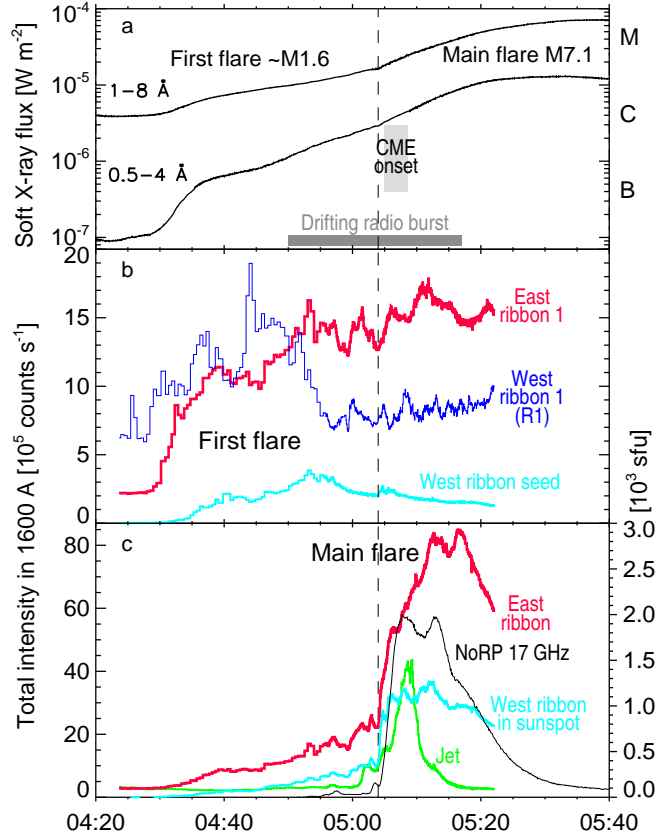}
  }
  \caption{Flare light curves recorded by GOES in soft X-rays (a) and
those computed from the TRACE 1600\,\AA\ images over the flaring
regions in the first flare (b) and main flare (c). The colors of
the curves correspond to those of the regions in
Figure~\ref{F-flare_regions}. Two distinct flare parts are
separated by the dashed line. The gray bars in panel a represent
the observation interval of a slowly drifting radio burst and the
CME onset time extrapolated to the position of AR~9742. The black
curve in panel c represents the background-subtracted total flux
at 17\,GHz. }
  \label{F-flare_light_curves}
  \end{figure}

In the main flare, the east ribbon broadened and extended
southwest (Figure~\ref{F-flare_regions}b). The west ribbon seed
broadened northwest and became the major west ribbon. A jet (green
contour) appeared after 05:06 in a funnel-like configuration with
a ring base, along which brightenings ran in the movie. Such
funnels appearing above magnetic islands inside opposite-polarity
regions contain coronal null points \citep{Masson2009,
Meshalkina2009}. The magnetic structure of a small flux-rope
erupting inside a funnel cannot survive at a null point
\citep{Uralov2014} and released plasma flows out as a jet
\citep{Filippov2009}. The collision of an erupting flux-rope with
a separatrix surface can produce a shock wave
\citep{Meshalkina2009, Grechnev2011}.

Figure~\ref{F-flare_light_curves} shows the flare light curves and
milestones of the event. An inflection in the GOES channels in
Figure~\ref{F-flare_light_curves}a separates the two parts of the
flare. The separation time is close to the CME onset time of
05:06\,--\,05:10 extrapolated to $1\mathrm{R}_\odot$ in the CME
catalog (\url{cdaw.gsfc.nasa.gov/CME_list/};
\citealp{Yashiro2004}) based on the data from the SOHO's
\textit{Large Angle and Spectroscopic Coronagraph} (LASCO:
\citealp{Brueckner1995}). The heliocentric distance of the flare
site was $0.864\mathrm{R}_\odot$. With an average CME speed of
1446\,km\,s$^{-1}$, its onset time at the flare site should be 65
seconds earlier (the light-gray bar in
Figure~\ref{F-flare_light_curves}a). The association of the CME
with the main flare was previously found by \cite{Gopalswamy2012}.

A slowly-drifting Type II and/or Type IV burst was associated with
the event. In the Learmonth spectrum up to 180~MHz shown by
\cite{Nitta2012}, the burst is detectable from about 04:57, too
early for the fast CME. The burst can be followed back until 04:50
up to 300~MHz in the \textit{Hiraiso Radio Spectrograph} (HiRAS)
spectrum (2001122605.gif) at
\url{sunbase.nict.go.jp/solar/denpa/hirasDB/Events/2001/} (the
dark-gray bar in Figure~\ref{F-flare_light_curves}a). This burst
could only be caused by expanding ejecta or a wave, which started
well before the fast CME from AR~9742. No other CME was detected
around that time. Most likely, this burst was due to a slower
eruption preceding the fast CME.

Figures \ref{F-flare_light_curves}b and
\ref{F-flare_light_curves}c present the light curves for the first
and main flare, respectively, calculated from the TRACE 1600\,\AA\
images as total over the major flare regions denoted in Figures
\ref{F-flare_regions}a and \ref{F-flare_regions}b. The bars
represent the imaging intervals. The black curve in
Figure~\ref{F-flare_light_curves}c shows a microwave burst
recorded by the \textit{Nobeyama Radio Polarimeters} (NoRP:
\citealp{Torii1979, Nakajima1985}) at 17\,GHz.

The flare has two distinct parts. The first flare started at about
04:30 and lasted $\approx 34$ minutes before the main flare. The
17\,GHz burst did not exceed 200\,sfu. The main flare started at
about 05:04 with a sharp increase of the UV emissions from the
both major ribbons and the 17\,GHz burst. The temporal profiles of
the west ribbon and microwave burst are similar but not identical.

\begin{figure} % {3}
  \centerline{\includegraphics[width=\textwidth]
   {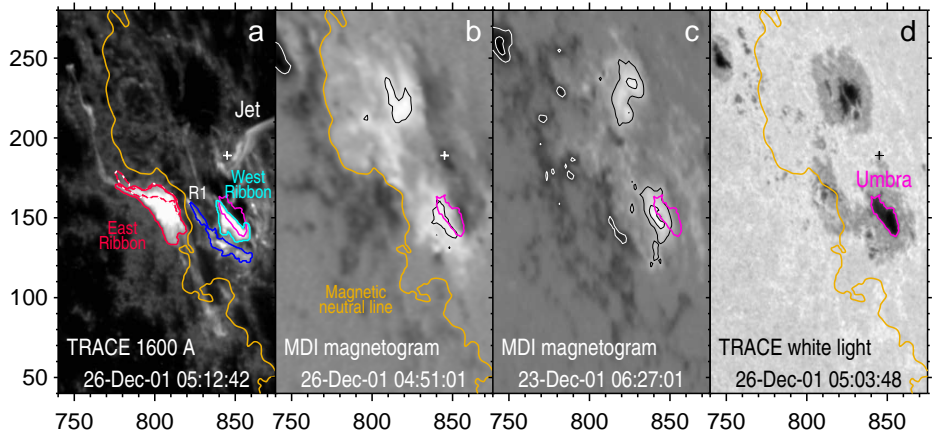}
   }
  \caption{Comparison of the flare configuration visible in a TRACE
1600\,\AA\ image (a) with the MDI magnetograms observed on 26
December (b) and 23 December (c) as well as a TRACE white-light
image (d), all transformed to their appearance at the onset time of
the main flare of 05:04. The broken red--white contour and the
dark-blue one outline the ribbons observed in the first flare. The
orange contour traces the magnetic neutral line. The pink contour
outlines the sunspot umbra. The contour levels for the magnetograms
are $[\pm 1000, \pm 2000]$\,G (black N-polarity, white S-polarity).
The crosses in panels a, b, and d mark the base of the jet.}
  \label{F-trace_mag_wl}
  \end{figure}

\subsection{Photospheric Configuration}

Figure~\ref{F-trace_mag_wl} compares the configuration observed in
1600\,\AA\ near the flare peak with MDI magnetograms and a
white-light TRACE image produced at the main flare onset. The
field of view presents the total length of the east ribbon with
its remote extensions into a northeast S-polarity sunspot and a
southwest region of weak magnetic fields. The color contours
correspond to Figure~\ref{F-flare_regions}. The red--white-dashed
contour denotes the east ribbon in the first flare. The pink
contour corresponds to the N-sunspot umbra in
Figure~\ref{F-trace_mag_wl}d. In the main flare, the west ribbon
overlapped with the sunspot umbra in Figure~\ref{F-trace_mag_wl}a,
as Article~I assumed.

The magnetogram of AR~9742 located not far from the limb is
affected by projection effects on magnetic field inclined to the
line of sight. To find realistic fields, we consider the
magnetograms observed just before the main flare
(Figure~\ref{F-trace_mag_wl}b) and three days before
(Figure~\ref{F-trace_mag_wl}c). The magnetograms are scaled by a
power-law function with a $\gamma$ of 0.8 for each polarity
separately (bright positive, dark negative).

AR~9742 evolved over three days. The major magnetic-field
distribution persisted. The changes are mostly related to
weaker-field regions and the shape of the neutral line. The
S-polarity field under the future east ribbon concentrated. The
flare-related south N-sunspot became slightly displaced.

The magnetograms exhibit three kinds of distortions. i)~Unlike the
magnetogram in Figure~\ref{F-trace_mag_wl}c, the west parts of the
sunspots in Figure~\ref{F-trace_mag_wl}b appear with a spurious
inverted polarity (this is a common feature in magnetograms close to
the limb). ii)~The field strengths in Figure~\ref{F-trace_mag_wl}b
are reduced \textit{vs.} Figure~\ref{F-trace_mag_wl}c both in
sunspots and plage regions, and a secant correction seems to be
justified for all of the regions. iii)~A hook-like shape of the
2000\,G contour in the south sunspot indicates a saturation-like
distortion occurring in MDI magnetograms. This distortion is also
present in the 26 December magnetogram. With a maximum field
strength of 2807\,G observed there on 23 December, the real strength
in the south sunspot could reach $\approx 3000$\,G \textit{vs.}
1164\,G in the 26 December magnetogram.

The cross in Figure~\ref{F-trace_mag_wl}a denotes the base center
of the funnel-like configuration, where the jet occurred.
Figure~\ref{F-trace_mag_wl}b shows there an island of an enhanced
magnetic field up to $-855$\,G. The magnetograms of the preceding
days indicate that, most likely, this was a negative-polarity
island inside a positive environment.

\begin{figure} % {4}
  \centerline{\includegraphics[width=0.85\textwidth]
   {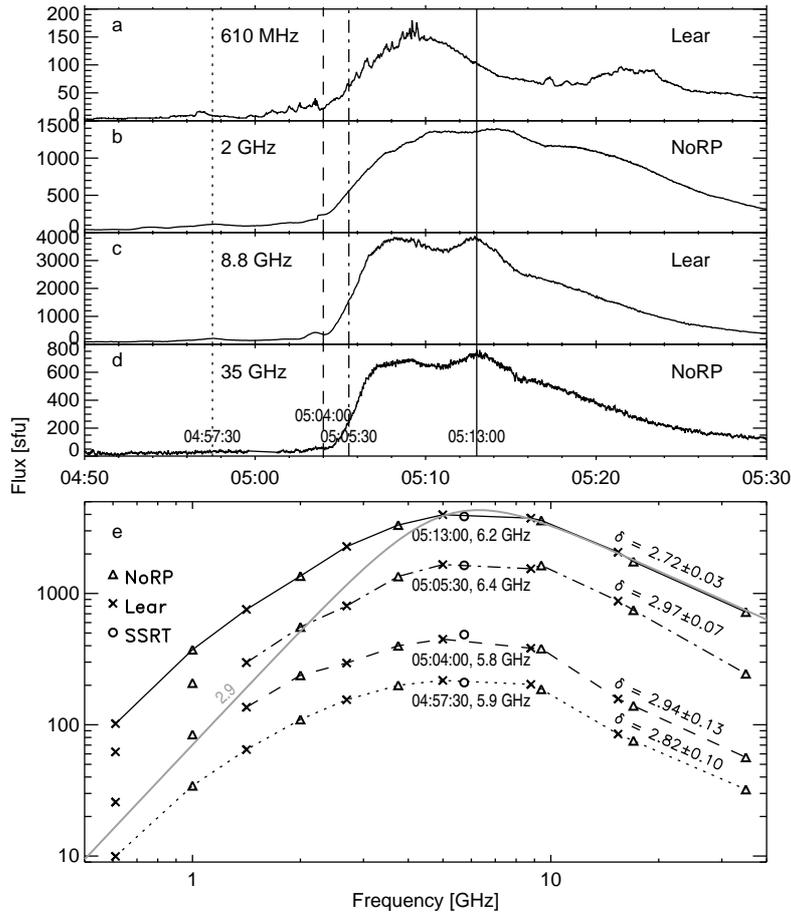}
  }
  \caption{a--d)~Total flux temporal profiles of the burst at four
radio frequencies. e)~The spectra for the four instances denoted
in panels a\,--\,d by the lines of the corresponding styles. The
observation time, estimated peak frequency [$\nu_\mathrm{peak}$],
and a power-law index of microwave-emitting electrons [$\delta$]
are specified for each spectrum. The gray curve represents a
classical GS spectrum corresponding to the high-frequency
observations at the peak of the burst.}
  \label{F-norp_lear}
  \end{figure}

\subsection{Microwave Observations}
 \label{S-microwave_observations}

The microwave burst was observed at a number of fixed frequencies by
the NoRP and Learmonth radiometers. These data allow us to analyze
the spectrum of the burst. We do not use the NoRP data at 80\,GHz,
which look unreliable. To cross-calibrate the NoRP and Learmonth
data, the background-subtracted flux at each frequency was
integrated from 04:20 to 05:30. The logarithmic spectrum was fitted
with a fifth-order polynomial, and the deviations from the fit were
used as cross-calibration coefficients, which ranged between 0.87
and 1.21.

Figures \ref{F-norp_lear}a\,--\,\ref{F-norp_lear}d present total
flux temporal profiles at four frequencies. The gyrosynchrotron
(GS) temporal profile extends to a very low frequency of 610\,MHz.
A spiky component is due to plasma emission, and late-phase
enhancements at 610\,MHz and 2\,GHz are probably due to a Type~IV
burst from a different source.

The spectra in Figure~\ref{F-norp_lear}e were calculated with an
integration time of 50 seconds for the four instances denoted in
the upper panels. The spectra are similar, although the ratio of
the peak fluxes between the solid curve and the dotted one reached
18. The peak frequency estimated from a sixth-order polynomial fit
was 6.1\,GHz\,$\pm\,5\,\%$. The low-frequency branch was elevated
relative to the slope of 2.9 for a classical GS spectrum (gray
curve) and had an actual slope of $\leq 1.8$ at frequencies $\geq
1.4$\,GHz. The elevation could be due to inhomogeneity of the
emitting source, \textit{i.e.} increasing at lower frequencies
contribution from higher layers, where magnetic field is weaker
(see, \textit{e.g.}, \citealp{LeeGaryZirin1994};
\citealp{Kundu2009} and references therein). The high-frequency
slope [$\alpha$] corresponds to a power-law index $\delta =
(1.22-\alpha)/0.9 \approx 3$ of the electron number spectrum,
according to \cite{DulkMarsh1982}.

\begin{figure} % {5}
  \centerline{\includegraphics[width=0.75\textwidth]
   {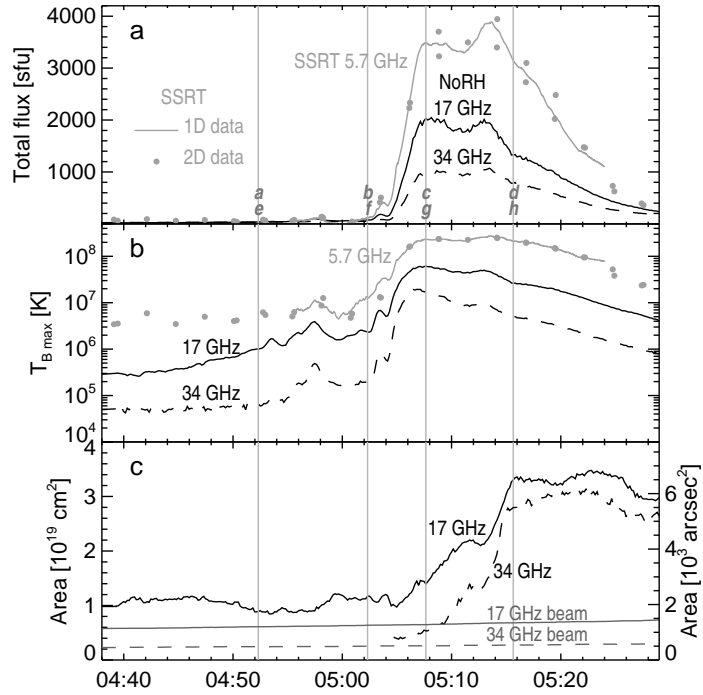}
  }
  \caption{Time profiles of microwave sources observed
at 17\,GHz (solid), 34\,GHz (dashed), and at 5.7\,GHz (gray):
a)~total flux, b)~maximum brightness temperature, c)~total area
(NoRH only). The nearly horizontal lines in panel c represent the
areas of the half-height NoRH beam at 17 and 34\,GHz. The labels at
the bottom of panel a denote the imaging times in
Figure~\ref{F-norh_trace}.}
  \label{F-norh_timeprof}
  \end{figure}

Figures \ref{F-norh_timeprof}a and \ref{F-norh_timeprof}b show the
total flux and maximum brightness temperatures of microwave
sources measured at 17 and 34\,GHz from NoRH images and at
5.7\,GHz from SSRT images (see Article~I).
Figure~\ref{F-norh_timeprof}c presents the total area of the
sources at 17 and 34\,GHz. The only way to increase the flux of a
simple GS source while keeping its spectrum shape is to increase
its area \citep{DulkMarsh1982, StahliGaryHurford1989} with an
unchanged brightness temperature. Figure~\ref{F-norh_timeprof}
presents an opposite relation. When the total flux and brightness
temperature strongly increased between the b and c instances, the
change in the total area of the sources was minor.
Figure~\ref{F-norh_timeprof}c also indicates that each of the
microwave sources might not be well resolved. This complicates the
situation.

\subsection{Motions of Microwave Sources and UV Ribbons}
 \label{S-motions_mw_ribbons}

\begin{figure} % {6}
  \centerline{\includegraphics[width=\textwidth]
   {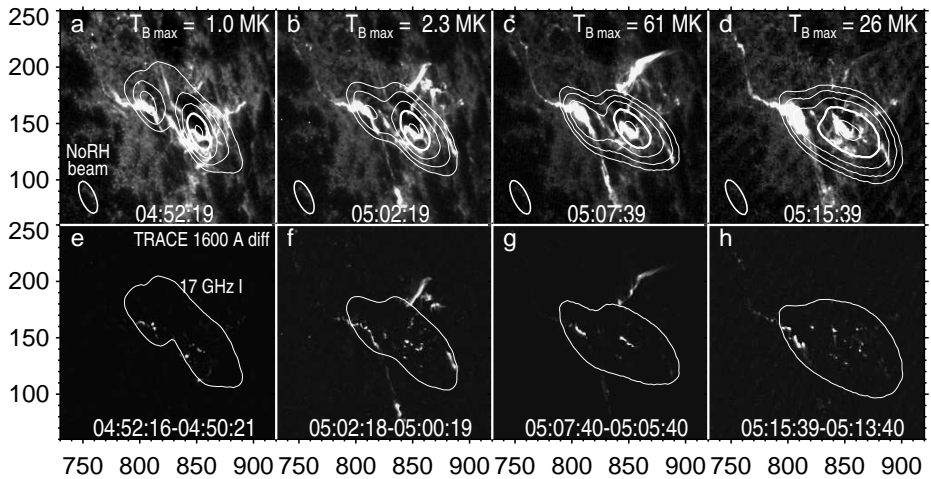}
  }
  \caption{Top: four flare episodes in 1600\,\AA\ (background) and
17\,GHz images (contours) denoted in Figure~\ref{F-norh_timeprof}.
Contour levels are at $[0.0625, 0.125, 0.25, 0.5,
0.9]T_{\mathrm{B}\, \max}$ specified in each panel. The thick
contour is at a half-height level. The ellipses represent the
half-height contours of the NoRH beam. Bottom: two-minute
running-difference 1600\,\AA\ images and 0.0625-level contours of
the 17\,GHz images. Solar rotation was compensated in all images
to 05:04.}
  \label{F-norh_trace}
  \end{figure}

To get indications of microwave-emitting structures, we consider
bright kernels in 1600\,\AA\ as footpoints of the loops, with whose
legs microwave sources could be associated.  Figures
\ref{F-norh_trace}a\,--\,\ref{F-norh_trace}d show the 1600\,\AA\
images overlaid by contours of the 17\,GHz images (see also the
\textsf{2001-12-26\_NoRH\_TRACE\_kernels.mpg} movie with contours at
$[0.1, 0.3, 0.9]T_{\mathrm{B}\, \max}$ over each 17\,GHz image).
Figures \ref{F-norh_trace}e\,--\,\ref{F-norh_trace}h reveal bright
kernels in 1600\,\AA\ emphasized by subtracting the images observed
two minutes before, along with the outermost 17\,GHz contours. With
a position of the flare site at N08\,W54, the microwave sources in
the low corona must be offset west-northwest from the
upper-chromosphere 1600\,\AA\ images. The offset uncertainty can
reach $10^{\prime \prime}$, while the coalignment accuracy is
presumably within $5^{\prime \prime}$ among the 17\,GHz images and
within $1^{\prime \prime}$ among the UV images.

Two microwave sources in Figures
\ref{F-norh_trace}a\,--\,\ref{F-norh_trace}d reside above two
ribbons located in opposite magnetic polarities. Such sources are
usually interpreted as the conjugate legs of a single loop. However,
different clusters of bright kernels in Figures
\ref{F-norh_trace}e\,--\,\ref{F-norh_trace}h correspond to different
sets of loops dominating at each of the four times. Comparison of
the thick contours and the NoRH beam (both at a half-height level)
shows that these sets of loops were unresolved by NoRH. If the two
17\,GHz sources were located in the legs of multiple loops, then
correspondence is expected between the relative positions of the
ribbons, on the one hand, and those of the microwave sources, on the
other hand.

\begin{figure} % {7}
  \centerline{\includegraphics[width=0.75\textwidth]
   {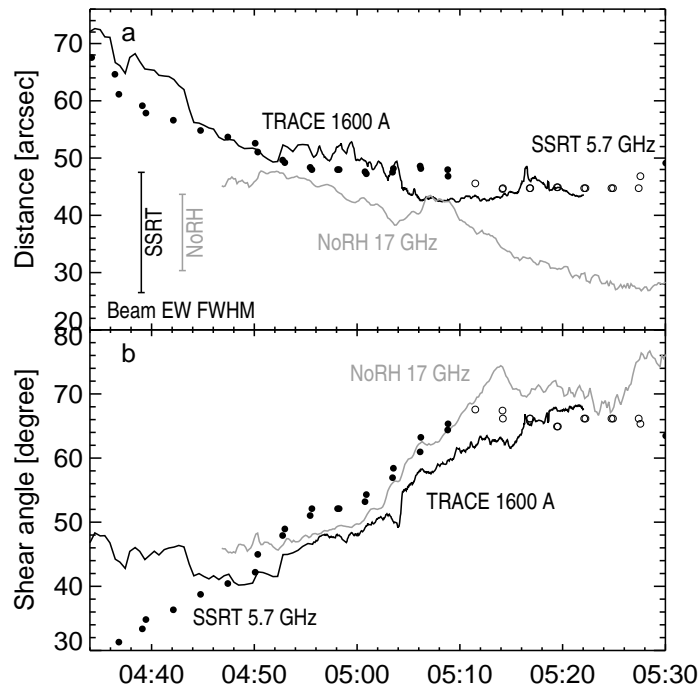}
  }
  \caption{a)~Temporal variations of the distance between the centers
of the two sources observed at 17\,GHz (NoRH: gray) and 5.7\,GHz
(SSRT: circles) in comparison with those for the flare ribbons in
1600\,\AA\ (TRACE, black). The vertical bars represent the full
widths at half maximums (FWHM) of the SSRT and NoRH beams in the
east--west direction. b)~The angle between the line connecting the
centers of the two sources and a major orientation of the magnetic
neutral line ($105^{\circ}$). The accuracy of the results denoted
by the open circles can be reduced due to overlap between the
images of the sources produced by the SSRT.}
  \label{F-kernels_timeprof}
  \end{figure}

Indications of this correspondence were found in Article~I from
the relative motions of the two sources observed in SSRT images at
5.7 GHz. Here we elaborate this result using the measurements from
the 1600\,\AA\ and 17\,GHz images. For the measurements at 17\,GHz
we used the same technique as in Article~I. For each ribbon we
used the centroid of its current image in 1600\,\AA\ within the
10\,\%-contour of the ribbon averaged over the whole flare.

Figure~\ref{F-kernels_timeprof} shows the distance between the
microwave sources and their position angle relative to the major
orientation of the neutral line ($105^{\circ}$ from the West)
measured from the 5.7 and 17\,GHz images and the same parameters of
the ribbons in 1600\,\AA. These relative measurements do not depend
on any coalignment accuracy. The east 17\,GHz source was absent
before 04:46:50, and our analysis of the TRACE data ends at
05:22:04. The vertical bars in Figure~\ref{F-kernels_timeprof}a show
that the variations of the distances are well under the beam size of
both SSRT and NoRH. This fact does not contradict the instrumental
resolution, because we measure the centroid of a source rather than
its structure, but it makes the results sensitive to imaging issues
and changes in the shapes of the sources.

\begin{figure} % {8}
  \centerline{\includegraphics[width=\textwidth]
   {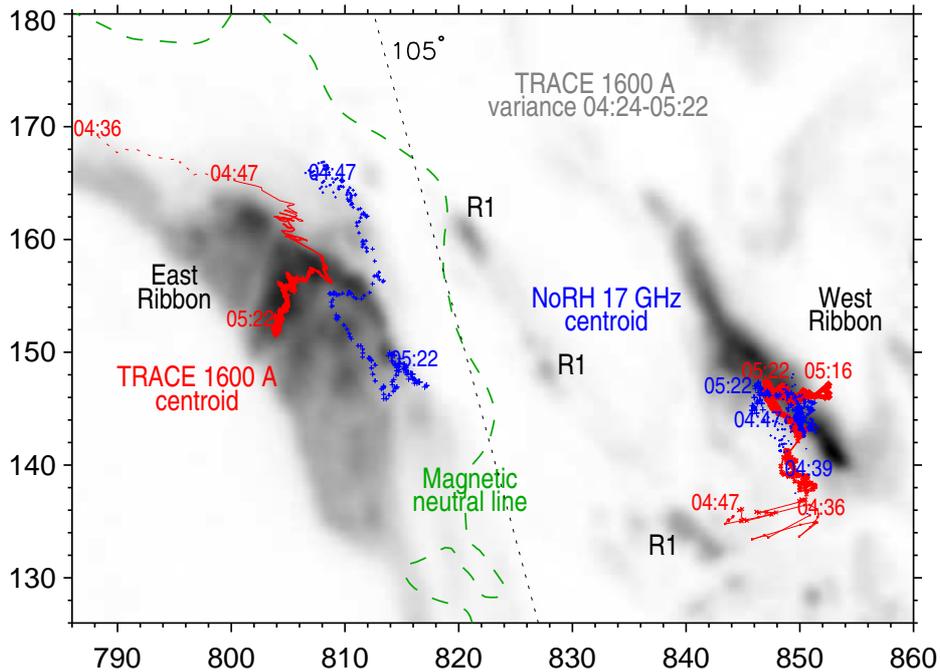}
  }
  \caption{Motions of of the flare ribbons (centroids) in 1600\,\AA\
images (red) and those of the 17\,GHz sources (blue). The gray scale
background presents a negative variance image computed over both
flare parts. Faintly visible parts of the west ribbon in the first
flare are denoted R1. The temporal succession is indicated by the
increasing thickness of the line (1600\,\AA) and size of the symbols
(17\,GHz). The green dashed line represents the magnetic neutral
line. The dotted line inclined by $105^{\circ}$ to the West is a
reference direction in the measurements of the shear angle in
Figure~\ref{F-kernels_timeprof}b. See also the
\textsf{2001-12-26\_NoRH\_TRACE\_kernels.mpg} movie.}
  \label{F-kernels}
  \end{figure}

The major tendencies observed in the three spectral ranges are
similar from 04:45 until 05:15. The microwave spectra in
Figure~\ref{F-norp_lear}e are also similar in this interval. The
decrease of the distance between the 17\,GHz sources in
Figure~\ref{F-kernels_timeprof}a after 05:10, dissimilar to the
others, is probably due to an increasing emission from entire low
loops filled with trapped higher-energy electrons. It is less
pronounced in the lower-frequency SSRT images and absent in the
1600\,\AA\ images. It is difficult to separate this contribution
from the main sources at 17\,GHz. The measurements from the SSRT
data denoted by the open circles are complicated due to overlap of
the sources (Article~I). The difference between the measurements
from the 5.7\,GHz and UV data in Figure~\ref{F-kernels_timeprof}b
before 04:45 is due to two causes. The first, geometrical cause is
the location of the UV ribbon centroid near a bend of the neutral
line (Figures \ref{F-flare_regions} and \ref{F-trace_mag_wl}) that
distorts the measured shear angle. Second, the magnetic field in
this region was insufficient to produce a detectable gyromagnetic
emission, and the 5.7\,GHz centroid was displaced to stronger
magnetic fields. With the complications mentioned, the relative
motions of the microwave sources and flare ribbons were similar.

This correspondence allows us to compare the measured centers of
the 17\,GHz sources and UV ribbons. We will not use here the
positions of microwave sources measured from SSRT images due to
their insufficient pointing accuracy. Figure~\ref{F-kernels}
presents the trajectories of the microwave sources (blue) and UV
ribbons (red). Their increasing thickness or size indicates time,
which is also specified at some positions. The gray-scale
background is a negative variance image computed from all TRACE
images obtained during the whole flare. This image represents all
changes observed in 1600\,\AA\ according to their statistical
contributions \citep{Grechnev2003varmap}. The green dashed line
represents the neutral line.

The variance image clearly shows the major east ribbon, which
extended south and broadened east, and the major west ribbon in
the main flare. The initial west ribbon in the first flare (R1)
appears in Figure~\ref{F-kernels} as three fragments.

The red-dotted line represents the centroid of the east ribbon
before 04:47, when its microwave counterpart was absent. The east
ribbon centroid moved nearly parallel to the neutral line from 04:36
to a hump, and then turned away from the neutral line. The of
trajectory the east microwave source is mostly parallel to that of
the east ribbon centroid. A later divergence is due to increasing
contribution to microwaves from the upper part of the arcade. The
offset of the microwave sources from the UV ribbons is due to their
different heights.

The west part of the flare site was more complex. Intermittent
brightenings on ribbon R1 during the first flare flipped the
centroid between R1 and west-ribbon seed. The west microwave source
was more stable due to a strong direct dependence on the magnetic
field. In the main flare, both the west ribbon and microwave source
drifted to the strongest-field region above the sunspot umbra. An
excursion of the ribbon centroid around 05:16 was due to a
brightening at the southwest edge of the sunspot that we do not
consider.

The agreement between the trajectories observed in 1600\,\AA\ and
at 17\,GHz cannot be a coalignment issue, because the west
microwave source was localized from 04:47 to 05:22 within
$6^{\prime \prime}$, while the east microwave source displaced
systematically during this interval by $14^{\prime \prime}$
(excluding the later mismatching part of its trajectory). The
trajectories of the ribbons and microwave sources correspond to
each other, except for the mentioned deviations, whose causes are
clear.

The measurements of the shear angle between the UV ribbons were
initially affected by a bending of the neutral line eastward. If
one mentally straightens this part of the neutral line and the
easternmost part of the east ribbon to keep the direction of
$105^{\circ}$ everywhere, then the initial angle would
substantially decrease. Thus, the overall decrease of the shear
throughout the event exceeded $40^{\circ}$ we measured.

\begin{figure} % {9}
  \centerline{\includegraphics[width=0.9\textwidth]
   {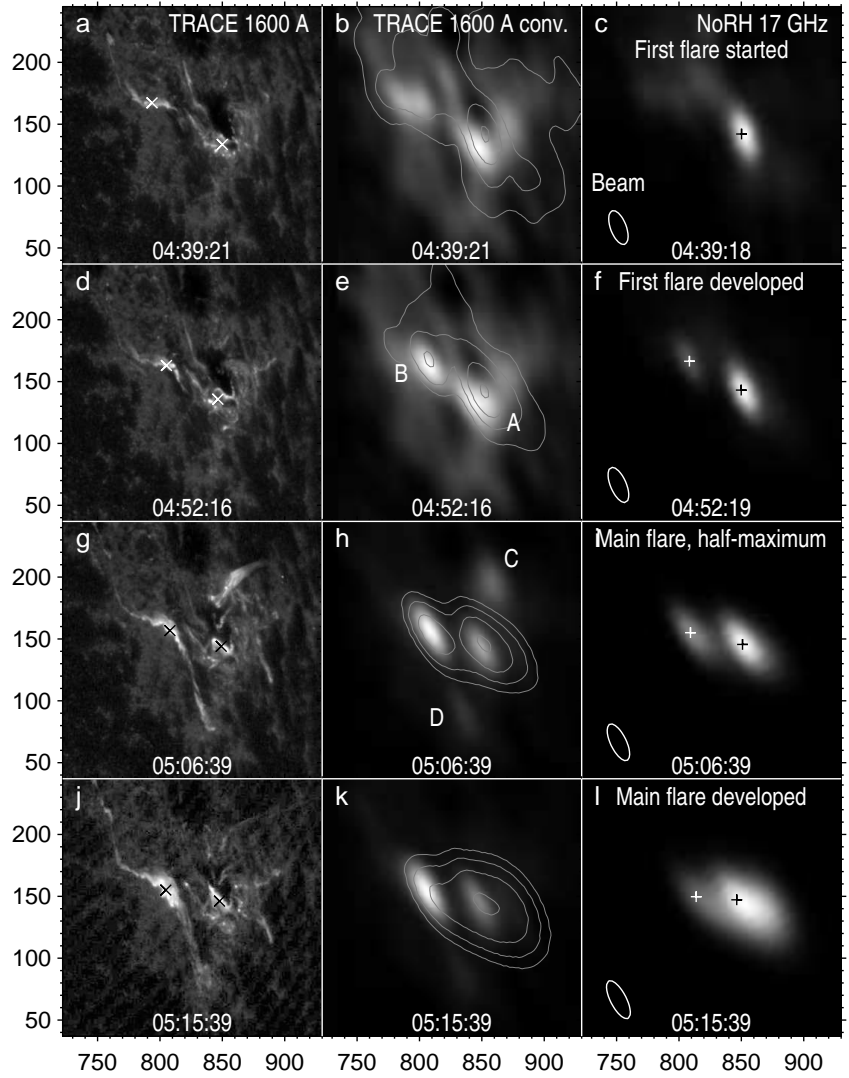}
  }
  \caption{Comparison of TRACE 1600\,\AA\ images (left column, logarithmic
brightness scale) with NoRH 17\,GHz images (right column, linear
scale). The middle column shows in the linear scale the 1600\,\AA\
images convolved with the NoRH beam (ellipses in the right panels)
overlaid with contours of the microwave images. Contour levels in
each image are at 0.9 of its maximum divided by powers of 3.
Centroid of each ribbon is denoted in the left column by the
slanted cross. Centroid of each microwave source is denoted in the
right column by the straight cross.}
  \label{F-trace_conv}
  \end{figure}

\subsection{Comparison of Microwave and UV Images}
 \label{S-comp_mw_uv}

The correspondence between the UV ribbons at the bases of the arcade
and microwave sources above the ribbons confirms association of the
microwave sources with the legs of the arcade loops. Non-thermal
electrons precipitating in dense layers are ultimately responsible
for the UV emission. On the other hand, both precipitating and
trapped non-thermal electrons radiate GS emission in the low-corona
magnetic loops. The UV and microwave emissions depend in opposite
senses on the magnetic-field strength [$B$]. With an estimated
electron spectrum index of $\delta \approx 3$, the GS emissivity at
optically thin frequencies has a direct dependence $\propto B^{2.5}$
\citep{DulkMarsh1982}, whereas the well-known mirroring in strong
magnetic fields hampers electron precipitation, which governs the UV
emission.

The real physical distinction between the structures emitting in
these two spectral ranges and responsible processes might be
emphasized by different instrumental characteristics of TRACE and
NoRH. To verify this idea, the difference in the spatial
resolution can be compensated by convolving the TRACE images
($1^{\prime \prime}$ resolution) with a NoRH beam. Its
cross-section is an ellipse, whose parameters gradually change
during the day. At 05:02 on 26 December, its half-height
dimensions were $12.7^{\prime \prime} \times 31.4^{\prime \prime}$
(we used the NoRH images synthesized at 17\,GHz by the
enhanced-resolution Fujiki software).

Figure~\ref{F-trace_conv} shows the results of our experiment for
four different flare episodes. The left column presents the TRACE
1600\,\AA\ images, the right column presents nearly simultaneous
NoRH images, and the middle column presents the TRACE images
convolved with the NoRH beam, whose elliptic half-height contours
are shown in the right panels. To facilitate comparison, the
convolved images in the middle column are overlaid with contours
of the microwave images.

The structures in Figures \ref{F-trace_conv}b and
\ref{F-trace_conv}c are similar. A subtle counterpart of the east
ribbon is also detectable at 17\,GHz. The environment in weaker
fields is faint but detectable in microwaves, as the contours in
Figure~\ref{F-trace_conv}b indicate.

The major sources in Figures \ref{F-trace_conv}e and
\ref{F-trace_conv}f are also similar. A brighter microwave source,
A, is located in stronger magnetic field. Conversely, the image of
the weaker-field east ribbon, B, is brighter in the convolved UV
image, as expected.

The presence of additional features C and D in
Figure~\ref{F-trace_conv}h makes it dissimilar to
Figure~\ref{F-trace_conv}i. Feature C is the jet emanating from a
configuration with a magnetic null point. The magnetic-field
strength there steeply falls off upward, and a microwave counterpart
is not expected. Feature D is a southernmost end of the east ribbon
located in a weak magnetic field. Its absence in microwaves is not
surprising, especially with a limited dynamic range of NoRH of about
300.

The images in Figures \ref{F-trace_conv}k and \ref{F-trace_conv}l
become less similar due to the upper part of the coronal arcade
appearing at 17\,GHz, but invisible in the UV. The brightness
temperature of the arcade in Figure~\ref{F-trace_conv}l exceeded
15~MK estimated for this time from GOES data in Article~I; the
power-law index of its brightness temperature spectrum estimated
from the images at 17 and 34\,GHz was around $-2.7$. Therefore,
thermal bremsstrahlung could only supply a minor contribution.
Most likely, the upper part of the arcade was dominated by trapped
electrons with a harder spectrum, consistent with a general
pattern established by \cite{KosugiDennisKai1988,
MelnikovMagun1998}, and in later studies.

In summary, the microwave sources A and B were, most likely,
associated with the legs of the arcade rooted in the ribbons. Bright
kernels in 1600\,\AA\ represented instantaneous loci of the electron
precipitation. Electrons trapped in numerous loops, whose footpoints
were shown by the UV kernels previously, emitted a prolonged
background GS emission. This long-lasting background reinforced the
similarity between the images of the long-lived flare ribbons and
microwave sources.

\subsection{Microwave Spectral Evolution}
 \label{S-spectral_evolution}

This possible scenario should be manifested in the evolution of
the microwave spectrum. In addition to the detailed GS spectra
presented in Figure~\ref{F-norp_lear}e for four instances, here we
examine the overall variations of the peak frequency
[$\nu_\mathrm{peak}$] and a power-law index of microwave-emitting
electrons [$\delta$] for the whole flare.

The $\delta$ index of the electron-number spectrum can be calculated
as $\delta = (1.22-\alpha)/0.9$ \citep{DulkMarsh1982}; $\alpha$
should be estimated from optically thin data, \textit{e.g.}, from
the total flux NoRP data at 17 and 35\,GHz or from NoRH images at 17
and 34\,GHz. We used both methods, because NoRP data are
characterized by a higher accuracy, while NoRH data provide a higher
sensitivity. The contribution of thermal bremsstrahlung was
subtracted. The result was smoothed over ten seconds.

\begin{figure} % {10}
  \centerline{\includegraphics[width=0.75\textwidth]
   {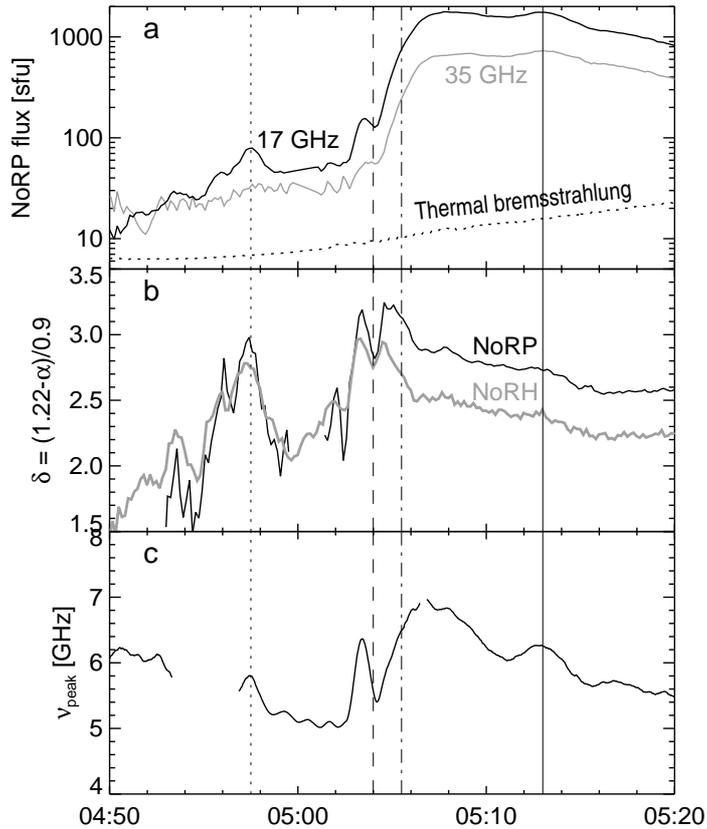}
  }
  \caption{Evolution of the microwave emission during the event.
a)~Total-flux temporal profiles recorded by NoRP at 17 and 35\,GHz.
The dotted line represents thermal bremsstrahlung estimated from
GOES data. b)~Power-law index of microwave-emitting electrons
computed from NoRP (black) and NoRH (gray) data. c)~Variations of
the microwave peak frequency. The vertical lines mark the four times
presented in Figure~\ref{F-norp_lear} with corresponding styles.}
  \label{F-delta}
  \end{figure}

The peak frequency was estimated in a way similar to the technique
used by \cite{White2003} and \cite{Grechnev2008, Grechnev2013a}
from NoRP and Learmonth total flux data
(Section~\ref{S-microwave_observations}). For each instance, a
combined spectrum at 12 frequencies was averaged over 16 seconds,
and a parabola was fitted to the five points of the log--log
spectrum closest to the peak. The result was smoothed over 30
seconds.

Figure~\ref{F-delta} shows the 17 and 35\,GHz fluxes along with an
estimated thermal flux, and the calculated $\nu_\mathrm{peak}$ and
$\delta$. The intervals with doubtful estimates are rejected. The
values of $\delta$ estimated from NoRH data are more reliable than
those from NoRP data for weaker fluxes and conversely for stronger
fluxes.

The electron index [$\delta$] in Figure~\ref{F-delta}b has an
impulsive component superposed on a harder gradual background. The
initial values of $\delta \approx 1.5-2.0$ are too hard, probably
due to underestimation of the thermal flux from the GOES data. They
are insensitive to plasma temperatures of $\lsim 3$~MK, whose
contribution to microwaves can be considerable. Its possible role
should lessen, as the GS emission increased. The GS emission from
trapped electrons with a progressively hardened spectrum in the
course of a continuous injection probably dominated late in the
event. The trapping effect could also affect the initial part of
Figure~\ref{F-delta}b, because the flare lasted about 20 minutes
before 04:50. Comparison of Figures \ref{F-delta}a and
\ref{F-delta}b shows that the impulsive component corresponds to
enhancements in the time-profiles. Most likely, freshly injected
electrons had a softer spectrum with $\delta \approx 3.5-4.0$, which
did not change considerably throughout the flare. A probable
power-law index of microwave-emitting electrons was mostly $\delta
\approx 2.5-3.3$.

The peak frequency in Figure~\ref{F-delta}c varies in a range of
5\,--\,7\,GHz. The values of $\nu_\mathrm{peak}$ (and $\delta$ in
Figure~\ref{F-delta}b) at the four marked instances are close to
those estimated in Figure~\ref{F-norp_lear}e in a slightly different
way. The evolution of $\nu_\mathrm{peak}$ suggests interplay of
parameters of emitting electrons and magnetic-field strength;
however, the range of $\nu_\mathrm{peak}$ seems to be atypically
narrow and low for the observed microwave fluxes. A possible key to
these challenges could be a distributed microwave-emitting system.
The average $\nu_\mathrm{peak}$ of individual sources was around
6\,GHz, and their total number elevated the peak flux of each one up
to the observed values. The scatter between the parameters of the
sources could be a reason for the broadening of the microwave
spectrum and its gradual shape.

\section{Discussion}
 \label{S-discussion}

Analysis of the microwave and UV images has shown that each of the
two microwave sources observed at 5.7, 17, and 34\,GHz was
associated with one of the two ribbons located in
opposite-polarity magnetic fields. The appearance of each
microwave source corresponds to the whole related ribbon, as if it
had been viewed by NoRH. Both the ribbons and microwave sources
exhibited nearly identical systematic motions. Until the peak of
the microwave burst, the two flare centers approached each other
nearly along the neutral line, so that the distance and shear
between them considerably decreased. After that, the motion
occurred away from the neutral line, corresponding to the usual
expansion of the ribbons. The expansion was measured from the SSRT
data until 06:30 in Article~I.

The correspondence of each microwave source to the flare ribbon,
their nearly identical motions, and surprisingly persistent shape
of the microwave spectrum, in spite of 18-fold flux variations,
indicate that microwaves were emitted by the conjugate legs of
multiple loops constituting the flare arcade. The relation of the
magnetic-field strengths under the ribbons points to a strongly
asymmetric configuration. Magnetic-flux conservation in flare
loops requires larger areas of the sources above the weaker-field
east ribbon relative to their conjugate counterparts above the
sunspot-associated west ribbon. To verify these considerations, we
will attempt to reproduce the observed spectra by means of a
simple model, using the magnetic fields actually measured.

\subsection{Reconnection Power and Flux of GS Emission}
 \label{S-reconnection_power}

The microwave flux density [$F(t)$] at optically thin frequencies
[$\nu>\nu_\mathrm{peak}$] is controlled by the instantaneous total
number of emitting high-energy electrons [$N_\mathrm{tot}(t)$].
Self-similarity of the spectra in a wide range of microwave fluxes
in Figure~\ref{F-norp_lear}e shows that the $N_\mathrm{tot}(t)
\propto F (t)$ relation in this event also applied at lower
frequencies [$\nu \leq \nu_\mathrm{peak}$]. Accelerated electrons
are produced in a reconnection process; therefore,
$N_\mathrm{tot}(t)$ is associated with a power of flare energy
release. Let us find which observable parameters of a flare indicate
this association.

The energy flux density entering the flare current sheet from one of
its sides is governed by the Poynting vector,
$\textbf{\textit{P}}=c[\textbf{\textit{E}}\times\textbf{\textit{B}}]/{4\pi}=
-[[\textbf{\textit{v}}\times\textbf{\textit{B}}]\times\textbf{\textit{B}}]/{4\pi}=
[\textbf{\textit{v}}B^{2}-\textbf{\textit{B}}(\textbf{\textit{B}}\cdot
\textbf{\textit{v}})]/{4\pi}=
{\textbf{\textit{v}}B^{2}}/{4\pi}$~[erg\,cm$^{-2}$\,s$^{-1}$]. Here
$\textbf{\textit{E}}=
-[\textbf{\textit{v}}\times\textbf{\textit{B}}]/c$;
$\textbf{\textit{B}}$ is a vertical magnetic field, \textit{i.e.}
$B=B_{z}$; $\textbf{\textit{v}}$ is a horizontal velocity of the
plasma inflow into the vertical current sheet, \textit{i.e.},
$v=v_{x}$.

The total power released in the current sheet dimensioned $Y$ by
$Z$ is $q = 2 P\, YZ =vB^{2}YZ/{2\pi}=BZ\, BY
(\mathrm{d}x/\mathrm{d}t)/{2\pi}=BZ\,
(\mathrm{d}\Psi/\mathrm{d}t)/{2\pi}$\,[erg\,s$^{-1}$]. Here
$\mathrm{d}\Psi=BY \mathrm{d}x$, and $\mathrm{d}\Psi/\mathrm{d}t$
is the input rate of the magnetic flux.

Let $\tau$ be the lifetime of a point-like UV kernel in the
footpoint of a thin magnetic tube during reconnection in the
current sheet and afterward. A multitude of kernels constitutes an
instant UV-emitting stripe corresponding to a narrow moving flare
ribbon in the standard model. Then, $\tau\,
\mathrm{d}\Psi/\mathrm{d}t =\Psi(\tau, t)$, which is magnetic flux
within the ribbon stripe at time [$t$]. Presumably, $\tau$ is
proportional to the lifetime of accelerated electrons in the
magnetic tube, and $\Psi(\tau, t)$ is proportional to the magnetic
flux across a GS source at time [$t$]. A particular value of
$\tau$ is not important, if it is much shorter than the burst.

If the flare process operated self-similarly throughout the burst,
then the ratio of energy released in the current sheet during
$\tau$, $W(\tau,t)=\int^{t+\tau}_{t}{q \mathrm{d}t} = q\tau = BZ\,
\Psi(\tau, t)/{2\pi}$, to the total energy of electrons produced at
the same time, $W^\mathrm{GS}(\tau,t)$, was constant, \textit{i.e.}
$W^\mathrm{GS}(\tau,t) \propto W(\tau,t)$. This relation is correct,
as long as the vertical size [$Z$] of the current sheet is constant,
and magnetic field [$B$] in its vicinity is uniform. The latter
assumption is justified by rapid disappearance with an increasing
height of small magnetic features, which reflect strong
inhomogeneity of the magnetic field on the photosphere.

With a power-law energy distribution of GS-emitting electrons
[$n(\epsilon) \mathrm{d} \epsilon = K\epsilon^{-\delta} \mathrm{d}
\epsilon$ ($\epsilon_{0} \leq \epsilon \leq \infty$)] their energy
density is ${\cal E}_{N} = [(\delta-1)/(\delta-2)] \epsilon_{0} N
\equiv \overline{\epsilon}N$\,[erg\,cm$^{-3}$], $\delta > 2$. Here
$n(\epsilon)$ is the number of electrons per cm$^{3}$ in a unit
interval of energy [$\epsilon$], $N=\int n(\epsilon) \mathrm{d}
\epsilon$ is the total number of GS-emitting electrons in
1\,cm$^{3}$, and $\overline{\epsilon}$ is an average energy of a
single electron.

Since $W^\mathrm{GS}(\tau,t) = N_\mathrm{tot}(t)
\overline{\epsilon}$, from the condition $ W^\mathrm{GS}(\tau,t)
\propto W(\tau,t)$ we get a final equation for the total number of
GS-emitting electrons, $N_\mathrm{tot}(t) = \mathrm{const}\times
BZ\Psi(\tau, t)/ \overline{\epsilon}\propto \Psi(\tau, t)$. Note
that $B$ is related here to the vicinity of the current sheet. In
turn, $\Psi(\tau, t)$ is an instant magnetic flux within one of
the ribbon stripes, $ \Psi(\tau, t) \equiv
\Psi^\mathrm{stripe}(t)$. The relation between the emission flux
and total number of emitting electrons, $F(t) \propto
N_\mathrm{tot}(t)$, is transformed to the form $F (t) \propto
\Psi^\mathrm{stripe}(t)$.

This relation, similar evolutions of microwave sources observed by
two radio heliographs and flare ribbons observed in UV, and
similarity between microwave images and convolved UV images
motivated our usage of a model source system referring to the
ribbons to simulate GS emission in this event.

\subsection{Modeling of Gyrosynchrotron Emission}

The spectrum of the non-thermal microwave emission in the 26
December 2001 event has two conspicuous features: persistent shape
with weak changes in the peak frequency under large flux
variations and an enhanced low-frequency part.
\cite{MelnikovGaryNita2008} found the peak-frequency variations to
be small in about one third of the events. The authors related
this behavior to GS self-absorption around the peak of the burst
and to the Tsytovich--Razin suppression in its early rise and late
decay. \cite{StahliGaryHurford1989} reported the latter effect
indeed; however, its importance at the rise of a long-duration
flare is difficult to reconcile with chromospheric evaporation. It
is quantified by the Neupert effect \citep{Neupert1968},
\textit{i.e.} similarity between the soft X-ray flux (directly
dependent on the plasma density) and antiderivative of the
microwave burst. The plasma density is initially low, reducing the
Tsytovich\,--\,Razin effect at this stage. At the decay of our
burst, \textit{e.g.} at 05:29, the net total area at 34\,GHz was
$A \approx 2.3 \times 10^{19}$\,cm$^2$
(Figure~\ref{F-norh_timeprof}c), emission measure estimated from
GOES data $\mathrm{EM} \approx 4 \times 10^{49}$\,cm$^{-3}$
(Article~I), and plasma density $\approx
(\mathrm{EM}/A^{3/2})^{1/2} \approx 1.9 \times
10^{10}$\,cm$^{-3}$. With a magnetic field strength of $B \approx
540$\,G estimated in Article~I at 05:20, the Razin frequency was
$\nu_\mathrm{R} = 2\nu_\mathrm{P}^2/(3\nu_B) \approx
0.68\,\mathrm{GHz} \ll \nu_\mathrm{peak} \approx 5.5$\,GHz
(Figure~\ref{F-delta}c). Thus, the ideas of
\cite{MelnikovGaryNita2008} are unlikely to help us, because the
magnetic fields of $\leq 300$\,G they considered are too weak for
our sunspot-associated flare.

To account for the low-frequency increase in a GS spectrum,
inhomogeneity of the source and superposition of multiple sources
have been proposed (\textit{e.g.} \citealp{Alissandrakis1984,
Alissandrakis1986, LeeGaryZirin1994, KuznetsovNitaFleishman2011}).
The major inhomogeneity in these models is related to the magnetic
field in a flare loop of a varying cross-section. This undoubted
inhomogeneity affects the shape of the spectrum, especially its
optically thick part \citep{BastianBenzGary1998, Kundu2009}. It is
difficult to understand why the spectrum from a single inhomogeneous
loop had a constant shape, while indications of multiple sources are
certain.

We are not aware of inhomogeneous multi-loop models. To verify our
interpretations, we are forced to use a tentative simplified
modeling of GS emission from a set of several homogeneous sources.
We are interested in general features of this system and need a very
simple analytic description of GS emission, which the
\cite{DulkMarsh1982} approximations present. Their reduced accuracy
at the lowest and highest harmonics of the gyrofrequency is not
crucial for the task of the model to understand properties of our
event.

Our model contains a considerable number of homogeneous GS sources,
each with a different magnetic-field strength and volume. Their
number depends on the width and length of the brightest parts of the
UV ribbons. The model should also demonstrate the direct dependence
of the total flux and spectrum of the microwave emission on the
total magnetic flux and its distribution over each of the ribbons.
The model does not consider the influence of the ambient plasma on
generation and propagation of the GS emission, \textit{i.e.} the
Tsytovich--Razin effect and free--free absorption, whose importance
in our event is unlikely.

The loop system constitutes an arcade rooted in the ribbons. Each
ribbon in our event is extended and inhomogeneous in brightness and
width. We relate a set of microwave sources to a brightest, broadest
stripe of each ribbon. Its width [$\Delta^{0}$] corresponds to a
typical transversal size of a loop, whose end is rooted in this
ribbon. The width of a loop varies according to the magnetic field
strength [$B$] along it, being equal to $\Delta^{0}
(B^{0}/B)^{1/2}$, with superscript ``0'' related to the first
ribbon. For a narrow ribbon stripe, the number of emitting loops
[$m$] should be about its length to width ratio. If the loops do not
overlap, then their total flux [$F(t)$] is the sum of the fluxes
emitted by all of the loops.

Each $i$th loop is represented by two homogeneous cubic sources in
its legs above both ribbons, corresponding to the observations at
5.7 and 17\,GHz. Magnetic fluxes in conjugate cubes are equal to
each other, $\Psi_{i}^\mathrm{E} = \Psi_{i}^\mathrm{W}$. The ratio
of their sizes [$l_{i}^\mathrm{E}/l_{i}^\mathrm{W}$] is determined
by the ratio of the magnetic-field strengths in the east source
[$B_{i}^\mathrm{E}$] and the west one [$B_{i}^\mathrm{W}$], so that
$\Psi_{i}^\mathrm{(E,\,W)} = B_{i}^\mathrm{E}{l_{i}^\mathrm{E}}^{2}
= B_{i}^\mathrm{W}{l_{i}^\mathrm{W}}^{2}$. It is convenient to use a
set of $m$ loops, each of which encloses equal magnetic flux
$\Psi_{i}= \Psi^\mathrm{stripe}/m$. This assumption ensures the
balance of magnetic fluxes in conjugate legs of any loop,
irrespective of its location, and facilitates partition into $m$
cubic sources. Two methods of partition are possible.

In the first method, the total magnetic flux
$\Psi^\mathrm{stripe}$ is divided into $m$ equal parts on the
magnetogram within each ribbon in a fixed direction. The widths of
the pieces can be different, while their magnetic fluxes are equal
to each other. Each $i$th pair corresponds to a loop. The loops do
not overlap, and the procedure to find $B_{i}$ and $l_{i}$ seems
to be physically justified.

A rather formal second method considers the histograms $\{B,\,
n(B)\}$ of the magnetic-field distribution within each ribbon,
where $n(B)$ is the number of pixels where the magnetic-field
strength is equal to $B$. The area under the histogram is divided
into $m$ equal parts corresponding to equal magnetic fluxes
$\Psi^\mathrm{stripe}/m$, which is easy to calculate. The cubic
sources obtained in this way are different, and their paired link
is lost. On the other hand, the scatter of the size and
magnetic-field strength is maintained, as in the first method of
partition. We use the second method, which is simpler to
implement.

The spectral flux density $F_i(\nu)$ from each $i$th unpolarized
source is $F_i(\nu)=2k T_{\mathrm{(eff)}
i}(\nu)\left[1-\exp{(-\tau_i(\nu))} \right] (\nu^2/c^2)\, A_i/R^2$,
where $k$ is Boltzmann's constant, $A_i$ the source area, $R =
1$~AU, and $\tau_i(\nu)= \kappa_i(\nu)l_{i}$ the optical thickness.
The effective temperature [$T_{\mathrm{(eff)} i}(\nu)$] and
absorption coefficient [$\kappa_i(\nu)$] are calculated following
\cite{DulkMarsh1982}. In their Figure~3, the log--log plots of
$T_{\mathrm{eff}}$ and $\kappa$ deviate at low $\nu/\nu_B$ from the
quasi-linear parts into opposite directions that reduces the errors
\citep{Kundu2009}. The deviations are less for $\delta < 3.5$ in our
case. The total flux spectrum is a sum of $2m$ spectra from all
sources. The number density of microwave-emitting electrons [$N$]
and their power-law index $\delta = (2.7-3.0)$ are identical for all
sources. The viewing angles of the sources in the legs of the loops
above the east and west ribbons $\theta^\mathrm{E}$ and
$\theta^\mathrm{W}$ are different, while their half-sum is the
longitude of the flare site.

The optimal number of the paired sources [$m$] was adjusted
iteratively to meet three conditions: i)~the sum of $2m$ sources
provides a gradual spectrum with a single peak, ii)~the value of $m$
is about the ribbon length to width ratio, and iii)~the model
acceptably fits the observed spectrum. With an optimal number [$m$],
the field strengths [$B_{i}^{(\mathrm{E, \, W})}$] estimated from
the photospheric magnetogram should be corrected to the coronal
values. It is possible to use a constant scaling factor $\mu$, so
that magnetic field strengths in coronal sources are $\mu
B_{i}^{(\mathrm{E, \, W})}$.

To estimate $\mu$, we refer to \cite{LeeNitaGary2009}, who found
an average magnetic field of $\overline{B} \approx 400$\,G using a
homogeneous GS source model and a scaling law between
$\overline{B}$ and the total area of a source. An intuitive option
to calculate $\mu$ as a ratio of 400\,G to an average field
strength measured from the magnetogram within the ribbons leads to
a biased estimate because of a nonlinear dependence of the
microwave flux on magnetic field.

With any number of the sources, the flux at optically thin
frequencies [$\tau(\nu)\ll 1$] is controlled by the total number
of emitting electrons and their emissivity. Thus, $F_i(\nu) =
\mathrm{const}\times N B_{i}^{\alpha}l_{i}^{3}\nu ^{1-\alpha}
(\sin\theta)^{0.65\delta-0.43} 10^{0.52\delta}$ with $\alpha = 0.9
\delta - 0.22$; $\theta$ is the viewing angle
\citep{DulkMarsh1982}. The constancy of the optically thin total
flux emitted by the same electron population with any number of
the sources, up to a single large one (subscript ``S''), results
in an equality $\sum N_i {l_i}^{3}{B_i}^{\alpha}=N_\mathrm{S}
{l_\mathrm{S}}^{3} {B_\mathrm{S}}^{\alpha}$. Using the constancy
of the total number of emitting electrons, $\sum N_i {l_i}^{3} =
{N_\mathrm{S}}{l_\mathrm{S}}^{3}$, we obtain an average
magnetic-field strength in an equivalent single source
$B_\mathrm{S}={\sum N_i {l_i}^{3}{B_i}^{\alpha}} / {\sum N_i
{l_i}^{3}}$. Finally, we have estimated $\overline{B} = \mu
B_\mathrm{S} \approx 400$\,G for the sources above each ribbon
separately and obtained $\mu \approx 0.56$. In this approach, the
magnetic flux [$\Psi_{i}$] retains, and the change from $B_{i}$ to
$\mu B_{i}$ results in a corresponding change in the size of each
$i$th source from $l_{i}$ to $l_{i}/\sqrt{\mu}$.

\begin{figure} % {11}
  \centerline{\includegraphics[width=0.65\textwidth]
   {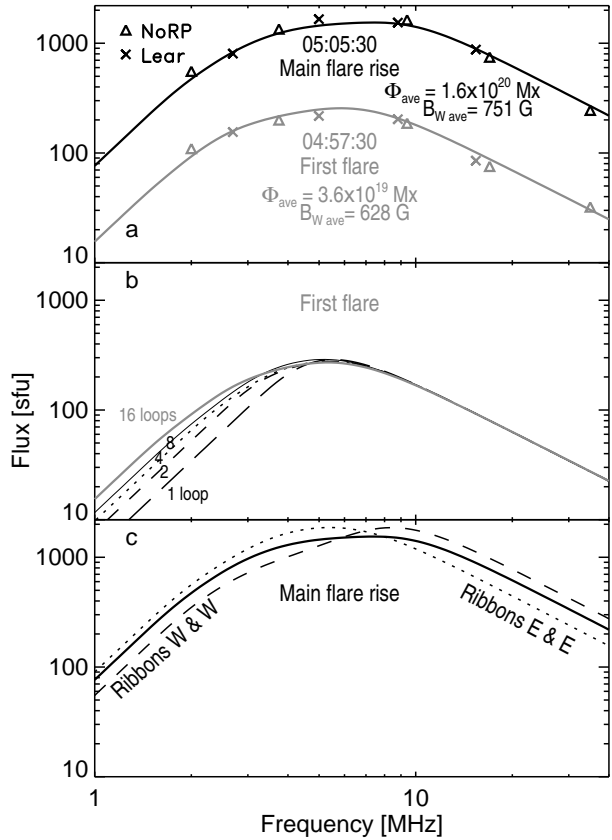}
  }
  \caption{Results of the modeling. a)~Observed (symbols) and modeled (lines)
GS spectra in the first flare (gray) and at the main flare rise
(black). b)~The spectra in the first flare modeled with a different
number of emitting loops from 1 to 16. c)~The influence on the
spectrum of asymmetry in the magnetic configuration: real
configuration (solid), symmetric configuration with a magnetic-field
distribution corresponding to the east ribbon at both sides
(dotted), and the same with that of the west ribbon at both sides
(dashed).}
  \label{F-model_spectrum}
  \end{figure}

Figure~\ref{F-model_spectrum}a presents the results of the
modeling by the gray line for the first flare (episode~1) and the
black line for the rise of the main flare (episode~2). These two
episodes were soon after fresh injections, when emission of
trapped electrons, which we do not consider, can be neglected, and
an assumption of a constant $\delta = 3.0$ is justified. The
symbols denote the observed fluxes. An enhancement at 2\,GHz was
due to a contribution from plasma emission. The average magnetic
flux over the ribbon stripe [$\Phi_\mathrm{ave}$] used in the
modeling is specified for each episode. The average field strength
above the west ribbon stripe [$B_\mathrm{W\, ave}$] additionally
affects the microwave flux at optically thin frequencies. The
number density of electrons with energies $> 10$\,keV was $N = 1.7
\times 10^{6}$\,cm$^{-3}$ in both episodes. The total number of
electrons was $N_\mathrm{tot\,1} = 1.9 \times 10^{33}$ in
episode~1 and $N_\mathrm{tot\,2} = 1.9 \times 10^{34}$ in
episode~2.

In Section~\ref{S-reconnection_power} we obtained $F(t) \propto
\Psi^\mathrm{stripe}(t)$ assuming the magnetic field to be uniform
at both sides of the neutral line that is not realistic.
Nevertheless, the model takes account of a real inhomogeneity of the
photospheric magnetic field and acceptably matches the real
spectrum. The relation between the magnetic flux and microwaves can
be generalized to a variable magnetic field, considering the
intrinsic dependence of the microwave flux on the magnetic field
strength, $F \propto N B^{\alpha}l_{S}^{3}= N_\mathrm{tot}
B^{\alpha}\propto \Psi^\mathrm{stripe}B^{\alpha}$.

The ratio of the optically thin microwave fluxes in the two
episodes in Figure~\ref{F-model_spectrum}a is 8.9, while the ratio
of the magnetic fluxes within the ribbon stripes is 4.4. With
$\alpha  = 0.9\delta-0.22 = 2.48$, the expected microwave flux
ratio is $4.4 \times (751/628)^{2.48} \approx 6.9$, which agrees
with the model result of 8.9 within $30\%$. This seems to be
acceptable with our simplified approach.

Figure~\ref{F-model_spectrum}b demonstrates the influence on the
spectrum of the number of loops for episode~1. The increase of the
number of loops elevates the low-frequency branch of the spectrum.
The effect is similar to that of the source inhomogeneity. The
ratio of the 16-loop spectrum to the single-loop one at 2\,GHz is
5.1 in this case. The lowest-frequency slope, 2.9, corresponds to
the classical GS spectrum.

Figure~\ref{F-model_spectrum}c illustrates the role of asymmetry
in the magnetic configuration with the same magnetic flux
reconnection rate. The solid line corresponds to the real
situation in episode~2. The dashed line represents the spectrum
for a hypothetical situation, when both microwave-emitting regions
were located above identical ribbons corresponding to the actual
west ribbon. The dotted line represents a similar experiment with
two east ribbons.

The strongest effect of asymmetry at optically thin frequencies is
illustrated by two extremities. The high-frequency flux from a
highly asymmetric configuration is determined by a single source,
and the flux is doubled in a symmetric configuration (two
identical sources). The effect of asymmetry varies between a
factor of one and two. The same occurs for the opposite asymmetry.
Thus, with the same magnetic-flux reconnection rate, the
high-frequency flux can vary within a factor of four.

The actual ratios of the optically thin fluxes in
Figure~\ref{F-model_spectrum}c are 1.26 between the dashed and
solid lines and 1.40 between the solid and dotted lines. The
asymmetry in this event progressively increased. The west ribbon
expanded into the sunspot umbra, the east ribbon developed into
weaker-field regions. To balance magnetic flux, an increasing
high-frequency emission from the stronger-field west regions must
be accompanied by an increasing area of the weaker-field east
region that elevated the low-frequency part of the spectrum. The
spectrum shape remained nearly constant, in spite of large changes
in the microwave flux.

The relevance of a homogeneous source and simplified
expressions by \cite{DulkMarsh1982} were discussed by
\cite{LeeNitaGary2009}. As they showed, the usage of a scaling law
between the average magnetic field and total source area makes the
homogeneous model sufficient to estimate statistical
characteristics of microwave bursts such as the peak flux and
frequency and spectral index. Each elementary loop in our model is
replaced by two homogeneous sources of different size and magnetic
field. A set of 32 homogeneous sources reflects inhomogeneity of
the ribbons. The scaling factor to shift from their magnetic
fields to those in microwave sources is based on the results of
\cite{LeeNitaGary2009}. Our model acceptably reproduces the
spectra also around the peak and at lower frequencies.

Modeling of the circular polarization of the GS emission
is complicated by the near-the-limb location of the flare site.
Since the west microwave source is visible through
quasi-transversal magnetic fields associated with the arcade,
polarization reversal is expected in a wide frequency range. Thus,
the polarization of each elementary source is a result of
interplay between the optical thickness and polarization reversal
issues, both frequency-dependent. This makes the analysis of the
polarization of the west source and the total emission too
complex.

The polarized emission of the east source can only be extracted in
NoRH 17\,GHz and SSRT 5.7\,GHz images. For the degree of
polarization of each optically thin elementary source we used a
corrected formula from \cite{Dulk1985}, otherwise we assigned an
opposite polarization of 15\%. The results for episode~1 are (all
negative; first observed, second model): $[25, 48]\%$ at 5.7\,GHz,
$[68, 54]\%$ at 17\,GHz; for episode~2: $[35, 55]\%$ at 5.7\,GHz,
$[67, 55]\%$ at 17\,GHz. The model reproduces the actual degree of
polarization within a factor of two with a correct sign.

The outcome of our simple modeling can be summarized as follows.

\begin{enumerate}

\item The observed properties of the GS spectrum are consistent
with the emission from a distributed multi-loop system,
\textit{i.e.} the flare arcade.

\item The asymmetry of the magnetic configuration is important.
The magnetic-flux balance requires larger areas at the weaker-field
side that elevated the low-frequency part of the spectrum, shifting
the peak frequency left. Inhomogeneities of the individual sources
could strengthen this effect.

\item To reproduce the observed spectrum with realistic magnetic
fields, a large increase of the magnetic flux is required in the
main flare (the average field strength changed insignificantly).
This result is consistent with a temporal correlation between the
magnetic-flux change rate and HXR emission found by
\cite{Asai2004, Miklenic2007, Miklenic2009}, who measured magnetic
flux within expanding parts of the ribbons.

\item Replacement of a distributed multi-loop system by a single
loop is generally not equivalent and can result in a different
shape of the spectrum, peak frequency, and their behaviors during
the burst. This leads to the next item.

\item Modeling of a microwave-emitting loop initiated by
\cite{Alissandrakis1984} has been developed into a powerful tool
\citep{TzatzakisNindosAlissandrakis2008,
KuznetsovNitaFleishman2011}. The models use real data on magnetic
field and take account of its inhomogeneity, anisotropy and spatial
distribution of electrons. A next-step challenge is a realistic
multi-loop model, at least, for simplified conditions. Elements of
the scheme presented here might be helpful in its development.

\end{enumerate}

\subsection{Motions of Flare Sources}

As known from observations in the H$\alpha$ line, flare kernels and
ribbons initially approach each other along the neutral line, and
then they move away from it. The expansion of the ribbons was
explained by the two-dimensional (2D) standard flare model. The
motions along the neutral line have not been clearly visualized.
Various motions observed later in HXR were summarized by
\cite{Bogachev2005}. The authors interpreted them in terms of the 2D
model; their cartoons implied a questionable rotation of the current
sheet around the vertical axis.

The motions along the neutral line have reasonably been considered
as an intrinsically 3D effect. Its scheme was presented by
\cite{Ji2008}. The authors supposed contraction of reconnected
loops to be a necessary element of the unshearing process, whereas
the two phenomena seem to be different results of the flare
reconnection not directly related with each other.

\cite{Reznikova2010} reported microwave observations of the motions
along the neutral line. The authors noticed that an M2.6 flare
developed along the arcade visible in the H$\alpha$ and extreme-UV
images and that, at least, several loops were involved in the
process. However, they considered a single loop for each instance.
In the discussion of the unshearing motions they mostly follow
\cite{Ji2008}.

\begin{figure} % {12}
  \centerline{\includegraphics[width=\textwidth,clip=,]
   {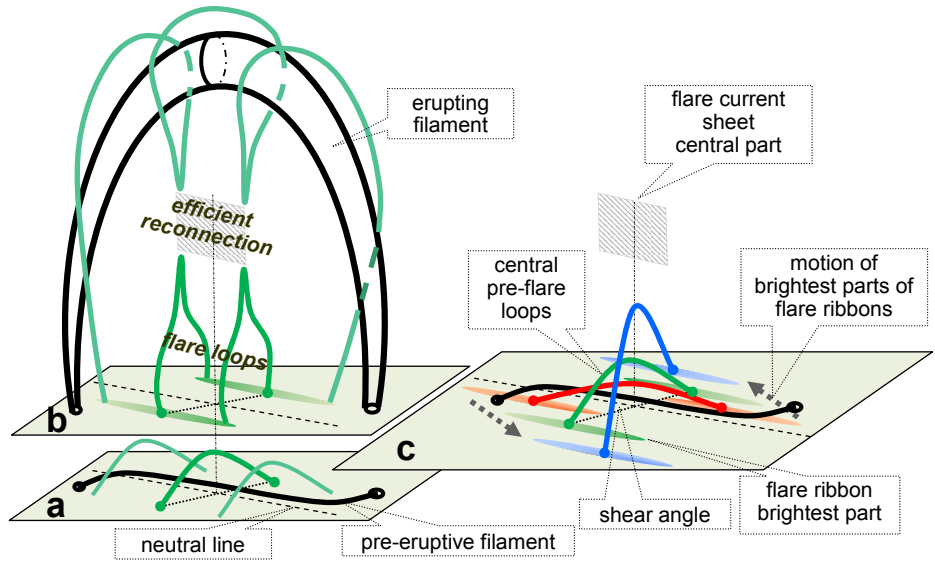}
  }
  \caption{Decrease of the shear between the brightest parts of
the ribbons in terms of a 3D flare model. a)~Before eruption. The
base corresponds to the photosphere. The filament (thick black) is a
flux-rope progenitor. Three coronal loops (green) belong to a single
magnetic surface. b)~Expansion of the filament and evolution of
magnetic field lines during eruption and flare. The brightest
segments of the ribbons (green stripes) correspond to a most
efficient central part of the current sheet. The centers of the
stripes coincide with the footpoints of the middle loop in panel a.
c)~Combination of panels a and b. The red, green, and blue loops
belong to different magnetic surfaces. The tops of these loops are
located under the central part of the future current sheet. Their
footpoints correspond to the brightest segments of the future
ribbons (red, green, and blue stripes). The shear between the
footpoints of the central loops decreases outward, from the red
stripe to the green one, and then to the blue one.}
  \label{F-scheme}
  \end{figure}

The motions of the microwave sources in our event corresponded to
those of the brightest parts of the ribbons, which we call the
ribbon stripes for brevity. Figure~\ref{F-scheme} explains the
relative displacement of the stripes. The intrinsic 3D geometry
implies the presence in an extended current sheet of a zone of
most efficient energy release, where the reconnection process is
most similar to the 2D model. The formation of major stripes is
associated with this zone. Figures \ref{F-scheme}a and
\ref{F-scheme}b present evolution of three magnetic loops, which
reconnected in this zone. Initially, these loops belonged to a
certain layer of magnetic arcades above the pre-eruptive filament.
The dotted line connects the bases of the central loop to show the
shear. Figure~\ref{F-scheme}c shows the central loops of different
layers. The higher a loop, the smaller is the shear of its bases.
The loops evolve similarly to the central loop in Figures
\ref{F-scheme}a and \ref{F-scheme}b. The major stripes
corresponding to these loops displace as the arrows show,
reflecting the decrease of the shear with the increasing height of
the pre-eruptive arcade.

The major motions initially occurred nearly along the neutral line
and away from it during the main flare phase and afterwards. This
behavior (not shown in Figure~\ref{F-scheme}) is associated with a
non-uniform decrease of the shear away from the neutral line. A
similar behavior is exhibited by helical lines of a cylindric
nonlinear force-free magnetic field ($\nabla \times
\textbf{\textit{B}} = \alpha\textbf{\textit{B}}$) when $\alpha \to
0$ with $r \to \infty$.

\subsection{Configurations Responsible for Thermal and Non-Thermal Emissions}

Non-thermal emissions are generated by energetic electrons and
carry the most direct information about acceleration processes in
flares. As noted in Section~\ref{S-introduction}, non-thermal
sources observed in HXR and microwaves are usually simple and
confined, which favors their identification with one or two
flaring loops. This view drawn from microwave observations of
almost all impulsive flares has been generalized to some major
long-duration flares \citep{TzatzakisNindosAlissandrakis2008}.
Still stronger confinement is typical of HXR sources; complex
structures like extended ribbons were observed in exceptional
events \citep{MasudaKosugiHudson2001, Metcalf2003,
LiuLeeGaryWang2007}. By contrast, observations of flares in
thermal emissions (H$\alpha$ line, UV, extreme UV, SXR) typically
show complex multi-component structures of a larger extent. This
dissimilarity has led to different views on flaring structures
drawn from different observations.

On the other hand, synergy between the structures observed in
thermal and non-thermal emissions has been conjectured previously,
including studies in which some of us participated
\citep{GrechnevNakajima2002, Grechnev2006acc, Kundu2009}. The causes
of the differences between non-thermal sources and configurations
visible in thermal emissions were unclear.

A key idea by \cite{MasudaKosugiHudson2001} explaining this
difference has not been commonly perceived. The authors pointed
out the limitations on the sensitivity and dynamic range
(typically $\approx 10$) in the HXR imaging. Indeed, images in HXR
and $\gamma$-rays have been provided by the imagers of the
Fourier-synthesis type with an intrinsically limited coverage of
the $(u,v)$-plane. These are the \textit{Hard X-ray Telescope}
(HXT: \citealp{Kosugi1991}) onboard \textit{Yohkoh} and the
\textit{Reuven Ramaty High-Energy Solar Spectroscopic Imager}
(RHESSI: \citealp{Lin2002, Hurford2002}). Sources weaker than
10\,\% of the brightest one are not detectable in their images
\citep{Krucker2014}. \cite{Asai2002} also found that HXR sources
accompanied the H$\alpha$ kernels in strongest magnetic fields
only; ``The HXR sources indicate where large energy release has
occurred, while the H$\alpha$ kernels show the precipitation sites
of nonthermal electrons with higher spatial resolution''.

Flares have been imaged in microwaves also almost exclusively with
Fourier-synthesis interferometers (mainly NoRH). They provide a
better coverage of the $(u,v)$-plane than HXR imagers. On the
other hand, optically thin GS emission strongly depends on the
magnetic field $\propto B^{\alpha}$ with $\alpha = 2.5-4$. With a
dynamic range of NoRH of $\approx 300$ \citep{Koshiishi1994}, its
opportunities to observe weaker non-thermal structures seem to be
comparable to HXR imagers. Also, microwave telescopes have a
poorer spatial resolution than HXR imagers.

For these reasons, the strongest non-thermal sources are only
expected in HXR and microwave images without weaker structures due
to instrumental limitations. Configurations, in which accelerated
electrons are manifested, and those visible in thermal emissions,
must actually be closely associated with each other.

Both thermal and non-thermal emissions in our event originated in
basically the same configuration. Dissimilarities between the
structures visible in microwaves and UV were mainly due to different
spatial resolution and dynamic range of the instruments and
different dependencies of the emissions on the magnetic field
strength. \cite{Zimovets2013} also concluded that, at least, some of
the single-loop configurations in NoRH images corresponded to
multi-loop arcades observed with telescopes of a higher spatial
resolution.

An appropriate proxy for a configuration responsible for
non-thermal emissions may be a structure observed in extreme
ultraviolet or in soft X-rays. This expectation corresponds to
widely accepted model concepts of eruptive flares (processes in
confined flares are unlikely different in nature -- see,
\textit{e.g.} \citealp{Thalmann2015}). This should be thoroughly
verified. If this is correct, then considerations of simple
configurations are justified, when they appear so in thermal
emissions. Inevitable simplifications should be recognized to
avoid inadequate conclusions.

\section{Summary and Concluding Remarks}
 \label{S-conclusion}

We have studied the 26 December 2001 eruptive flare, combining the
TRACE 1600\,\AA\ and NoRH 17 and 34\,GHz images, the results
obtained in Article~I from the SSRT 5.7\,GHz images, and different
data. The analysis has shown that the first flare and main flare
were most likely associated with separate eruptions.

\subsection{Milestones of the Event}

The first eruption presumably occurred in AR~9742 around 04:40 and
produced ejecta, which did not appear in the LASCO field of view
but manifested in a slowly drifting radio burst. A related
moderate two-ribbon flare involved medium magnetic fields,
produced a moderate microwave burst up to a few hundreds sfu at
5--6\,GHz, and reached a GOES importance of M1.6. This flare
lasted half an hour and had not fully decayed when another
eruption occurred in AR~9742.

The second eruption occurred around 05:04 and produced a fast CME.
The flare passed into the major two-ribbon flare and reached a
strength of M7.1. The east ribbon observed in the first flare
lengthened and broadened farther into regions of moderate and weak
magnetic fields. The west ribbon entered the strongest magnetic
fields above the sunspot umbra. The flare magnetic configuration was
increasingly asymmetric. The microwave burst strengthened up to
4000\,sfu at 6--7\,GHz and 780\,sfu at 35\,GHz, and lasted about 15
minutes (FWHM).

Furthermore, TRACE images reveal a jet-like eruption around 05:09.
Its light curve in 1600\,\AA\ is a spike with a FWHM duration as
short as three minutes. The jet is not detectable in microwaves
and will be analyzed in Article~III.

The first flare and the following main one were most likely caused
by the first and second eruptions, respectively. The eruptions
stretched closed magnetic configurations and thus could facilitate
escape of particles accelerated in the active region. Sharp
eruptions might have produced shock waves, which also could
accelerate heavy particles. Article~III will consider these
possibilities.

\subsection{Flare Morphology, Microwave Burst, and Proton Outcome}

A conspicuous morphologic manifestation of a large particle event is
flaring above the sunspot umbra \citep{Grechnev2013b}. This feature
indicates involvement in flare processes of the strongest magnetic
fluxes, whose reconnection rate corresponds to flare energy release
and governs particle acceleration. The flare ribbons in the events
analyzed previously overlapped with the umbras of opposite-polarity
sunspots. The SOL2005-01-20 (GLE69) and SOL2006-12-13 (GLE70) events
studied in detail exhibited large variations of the peak frequency,
whose maximum exceeded 25\,GHz, and peak fluxes at 35\,GHz exceeded
$10^4$\,sfu. While the major flare in the SOL2001-12-26 event looks
similar to these events, its morphologic difference is involvement
in the flare of a single sunspot.

The microwave flux directly reflects the magnetic flux
reconnection rate, being proportional to its instant value
multiplied by a factor of $B^{0.9\delta-0.22}$ at optically thin
frequencies. The major difference between the moderate first flare
and much stronger main flare was in the reconnected magnetic flux,
while parameters of the acceleration process manifesting in the
number density and power-law index of accelerated electrons
remained almost unchanged.

With the same magnetic-flux reconnection rate (and presumably the
same production of accelerated particles), the microwave response
strongly depends on magnetic fields, including symmetry of the
configuration. If it is asymmetric, then the microwave spectrum is
broader, the peak frequency is lower, and its variations are small.
The asymmetry of the configurations can cause an additional scatter
within a factor of four in the correlations between high-frequency
microwave bursts and near-Earth proton fluxes. An indication of
asymmetry is overlap of the flare ribbon(s) with the umbra of a
single, or two, or no sunspot.

Both flare parts were typical arcade flares, whose development is
described by the 3D model. Its intrinsic phenomena are the motions
of the arcade legs visible in microwaves with their bases visible
as the ribbons. Their initial approach nearly along the neutral
line reflects consecutive involvement in reconnection of
structures corresponding to the pre-eruptive magnetic field vector
distribution. It is similar to that one in a nonlinear force-free
flux rope, \textit{i.e.} from strongly sheared low structures to
less sheared those located at increasing distances from the axis.
A later expansion of the ribbons is accounted for by the 2D model.

In spite of a seeming single-loop configuration visible in microwave
images, the correspondence between the positions and motions of the
UV ribbons and non-thermal microwave sources evidences that
accelerated electrons emitted microwaves from the multi-loop arcade
rooted in the UV ribbons. This conclusion is supported by a simple
modeling of the microwave spectrum. Configurations with more than
one loop appear to be common in various flares: from small, spiky
events \citep{Kundu2004} to large, long-duration events
(\citealp{Grechnev2013a} and the present article). Our analysis has
demonstrated that dissimilarities between the structures visible in
non-thermal and thermal emissions are due to different instrumental
characteristics and different dependencies of the emissions on the
magnetic field. In accordance with well-known models, a proxy of a
configuration responsible for non-thermal emissions could be a
structure observed in thermal emissions.

\begin{acks}

We thank N.V.~Nitta for discussions and the reviewer for useful
remarks. We thank the instrumental teams managing SSRT, NoRH, and
NoRP; TRACE (NASA); SOHO/MDI and LASCO (ESA and NASA), GOES; USAF
RSTN Network; NICT (Japan); and the CME Catalog at the CDAW Data
Center (NASA and Catholic University of America). We appreciate
the memories of T.A.~Treskov, one of the major developers of the
SSRT, and N.N.~Kardapolova, who managed SSRT observations for many
years. This study was supported by the Russian State Contract
No.~II.16.1.6. A.~Kochanov was supported by the Russian Foundation
of Basic Research under grants 15-32-20504 mol-a-ved and
15-02-03717. V.~Kiselev was supported by the Marie Curie
PIRSES-GA-2011-295272 RadioSun project.

\end{acks}

\section*{Disclosure of Potential Conflicts of Interest} The authors
claim that they have no conflicts of interest.

\end{article}

\end{document}